\documentclass[useAMS,usenatbib,preprint]{mn2e}
\usepackage{graphicx}
\usepackage{color}
\usepackage{aas_macros}
\usepackage{gensymb}
\usepackage[T1]{fontenc}
\usepackage{float}

\usepackage[dvipsnames]{xcolor}
\definecolor{PColour}{rgb}{0.1,0.7,0.1}
\definecolor{IColour}{rgb}{0.65,0.,0.5}
\definecolor{RColour}{rgb}{0.1,0.2,0.9}
\definecolor{CColour}{rgb}{0.66,0.17,0.17}

\title[The dipole anisotropy of AllWISE galaxies]
{The dipole anisotropy of AllWISE galaxies}

\author[M. Rameez, R. Mohayaee, S. Sarkar, J. Colin]
{
M.~Rameez$^1$,
R.~Mohayaee$^{1,2}$,
S.~Sarkar$^{1,3}$,
J.~Colin$^{1,2}$
\\
$^{1}$ Niels Bohr Institute, University of Copenhagen, Blegdamsvej 17, 2100 Copenhagen {\O}, Denmark\\
$^{2}$ Sorbonne Universit\'es, UPMC Univ Paris 06, CNRS, Institut d'Astrophysique de Paris, 98bis Bld Arago, Paris 75014,France\\
$^{3}$ Rudolf Peierls Centre for Theoretical Physics, University of Oxford, 1 Keble Road, Oxford, OX1 3NP, United Kingdom
}

\date{\today}

\begin{document}
\maketitle
\label{firstpage}

\begin{abstract}
We determine the dipole in the WISE galaxy catalogue. After reducing star contamination to <0.1\% by rejecting sources with high apparent motion and those close to the Galactic plane, we eliminate low redshift sources to suppress the non-kinematic, clustering dipole. We remove sources within $\pm 5^0$ of the super-galactic plane, as well as those within $1''$ of 2MRS sources at redshift $z < 0.03$. We enforce cuts on the source angular extent to preferentially select distant ones. As we progress along these steps, the dipole converges in direction to within $5^0$ of the CMB dipole and its magnitude also progressively reduces but stabilises at $\sim0.012$, corresponding to a velocity >1000 km/s if it is solely of kinematic origin. However, previous studies have shown that only $\sim70\%$ of the velocity of the Local Group as inferred from the CMB dipole is due to sources at $z < 0.03$. We examine the Dark Sky simulations to quantify the prevalence of such environments and find that <2.1\% of Milky Way-like observers in a $\Lambda$CDM universe should observe the bulk flow (> 240 km/s extending to $z > 0.03$) that we do. We construct mock catalogues in the neighbourhood of such peculiar observers in order to mimic our final galaxy selection and quantify the residual clustering dipole. After subtracting this the remaining dipole is $0.0048 \pm 0.0022$, corresponding to a velocity of $420 \pm 213$ km/s which is consistent with the CMB. However the sources (at $z > 0.03$) of such a large clustering dipole remain to be identified.
\end{abstract}


\section{Introduction}
\label{sec:introduction}

The standard cosmological model assumes that the Universe is statistically isotropic and homogeneous on large scales. The isotropy of the Universe has supposedly been observationally confirmed by the Cosmic Microwave Background (CMB) temperature fluctuations on small angular scales which do not show any significant directional dependence \citep{wmap11,planck16}. However the dipole anisotropy of the CMB is $\sim 100$ times larger than that of the higher multipoles (see {\it e.g.} \cite{kogut93}). The latter are believed to have originated from an inflationary era in the early universe. Hence the CMB dipole is considered \emph{not} to be of primordial origin and is attributed to the motion of the Solar system in the `CMB rest frame' (in which the universe is exactly isotropic), due to the attraction of a nearby large density inhomogeneity. 

If the dipole anisotropy is indeed due to our motion, then the barycentre of the Solar system is moving at 369 km\,s$^{-1}$ towards RA=168$\degree$, DEC=$-7\degree$, or $l=263.85\degree$, $b=48.25\degree$ in Galactic coordinates \citep{stewart67, peebles68, hinshaw09}. Due to the indirect nature of this inference, which relies on there being no primordial dipole, an \emph{independent} direct measurement of this velocity is desirable. This can be done by observing the aberration of the CMB \citep{challinor02,burles06}, however the effect is too small to have been detected convincingly (with $>3\sigma$ significance) even using the latest data \citep{planck14}.

The standard $\Lambda$CDM model extended to first-order in perturbation theory does predict some anisotropy in the local Universe on small scales. These are however expected to become progressively smaller as the average is taken over larger volumes, leading to the emergence of homogeneity (and isotropy) on large scale. That this indeed happens on scales exceeding $\sim 100$ Mpc  has been claimed from observing the scale-dependence of counts of galaxies in the SDSS \citep{hogg05} and WiggleZ \citep{wigglez12} surveys. As the CMB is an integrated map of the Universe it cannot trace this transition, however galaxy surveys can indeed do so. They show that along the direction of the CMB dipole lie the most massive neighbouring superclusters: Virgo, the Great Attractor Hydra-Centaurus, Coma, Hercules and Shapley, and possibly other yet-to-be-mapped superclusters. However the gravitational attraction of these structures can account at most for 80\% of the velocity interpreted from the CMB dipole aniostropy \citep{lavaux10,colin11,feindt13}.

A dipole in the distribution of galaxies, which is usually known as the `clustering dipole' (${\cal D}_{\rm cls}$), can cause us to move in a preferred direction. This motion would lead to an additional anisotropy observable through the aberration and Doppler boost effects by an amount which depends on our velocity, further increasing the total observed dipole. A measurement of this latter effect, often referred to as the `kinematic dipole' (${\cal D}_{\rm kin}$), can provide an independent confirmation of our velocity. Aberration and Doppler boost effects can be measured only in a sample that is intrinsically isotropic, such as a directionally unbiased catalogue of high redshift sources. More often the total dipole (${\cal D}$) will be a mixture of the clustering and the kinematic dipoles. A further complication is  the dipole generated by statistical noise due to the finite size of the galaxy catalogue  (see {\it e.g.} \cite{Itoh10}), as well as contamination of the catalogue by foreground stars in our Galaxy. 

Most galaxy surveys are shallow and contain significantly large clustering dipoles from sources at distances less than a few hundred Mpc. Estimating our velocity  from these catalogues would require the disentangling of the clustering and the kinematic dipoles. Ideally surveys that are very deep in redshift should be able to provide observations at much larger scales, where the clustering dipole is expected to be subdominant. Radio galaxies are expected to provide such an alternative as they are extremely luminous and unaffected by obscuration due to dust, hence probe the Universe at high redshifts $z > 1$. Nearly all measurements of the kinematic dipole using radio galaxy catalogues are however \emph{discrepant} with the CMB dipole. While the velocity of the Solar system barycentre inferred from the CMB temperature dipole anisotropy is 369 km\,s$^{-1}$, the value inferred from radio-galaxy catalogues, {\it e.g.} NVSS, ranges from 700 km\,s$^{-1}$ to over 2000 km~s$^{-1} $ \citep{blake02,singal11,gibelyou12,rubart13,tiwari15,tiwarijain15,colin17}. The direction of the anomalously high velocity is however found to be quite well-aligned with the CMB dipole. 

Several previous studies have tested for homogeneity and isotropy in galaxy surveys with photometric redshifts e.g. the Sloan Digital Sky Survey (SDSS) Data Release 6 \citep{Itoh10}, the Two Micron All Sky Survey Photometric Redshift (2MRS) survey \citep{Appleby14,Alonso15}, the Wide Infrared Satellite Explorer (WISE) survey \citep{Yoon14}, the WISE-2MASS catalogue \citep{Bengaly17a}, and the WISExSUPERCOSMOS catalogue \citep{Bengaly17b}. We go beyond all these works by examining the \emph{largest} unbiased all-sky survey of the infrared sky, namely the Wide-field Infrared Survey Explorer --- All Sky (WISE-allsky) and its subsequent extension --- the AllWISE catalogues. As the WISE-allsky and AllWISE catalogues also contain stars and other point-like objects from within the Galaxy, these have to be first removed. Star contamination can be suppressed to the level of a few percent using methods described in \citet{szapudi15}. It can be further reduced to $<0.05\%$ by exploiting information from the apparent motion fits made possible by the NEOWISE post cryogenic phase of the WISE survey, supplied with the AllWISE catalogue. The Sloan Digital Sky Survey (SDSS)~\citep{SDSS} Data Release 13 is used as a reference catalogue to estimate star contamination through cross-correlation. 

In order to suppress the clustering dipole, we proceed to remove as many local sources as possible in a \emph{directionally unbiased} manner. This is done by first removing the sources that correlate with the 2MRS catalogue, as well as removing sources in symmetric bands around the supergalactic plane along which the most important superclusters in the local Universe lie. The sample can be further reduced by selecting for more distant galaxies by removing extended sources. At each of these steps, we evaluate the dipole. We show that as the clustering dipole is progressively suppressed, the total strength of the dipole is reduced as expected, while the direction converges towards the direction of the CMB dipole. We further test the robustness of this dipole by removing the well-known local superclusters (symmetrically, to avoid producing spurious dipole effects) such as Shapley directly from the catalogue. The Galaxy and Mass Assembly (GAMA)\citep{GAMA} is used as a reference spectroscopic survey to estimate the redshift distribution. 

At this stage, for a typical observer in a $\Lambda$CDM universe in a Milky Way-like halo ({\it i.e.} with similar mass and velocity), the remaining dipole should be dominated by the kinematical component. However the dipole we find at this stage is $\sim 0.0124$, implying a Solar system barycentre velocity of $\sim$1039 km\,s$^{-1}$ if it is indeed of purely kinematic origin. This is nearly three times as large as the velocity inferred from the CMB dipole.

In order to determine whether this final kinematic dipole is contaminated by any residual clustering dipole, we produce catalogues with the characteristics of our AllWISE selection from the $z=0$ halo catalogue of the Dark Sky simulation \citep{darksky}. We construct the catalogue slice by slice in comoving distance to reproduce the redshift distribution of our final selection. The catalogues are constructed around halos similar in mass and peculiar velocity of the Galaxy. In addition, we also consider constrained observers similar to us, {\it i.e.} in environments where the $z=0.03$ sphere around them has a bulk motion of 220--260 km\,s$^{-1}$. We find that the residual clustering dipole is \emph{larger} around such constrained observers, who constitute however $<3\%$ of Milky Way-like observers. After subtracting this estimate of the average expected clustering dipole from the total observed dipole, we find the remaining dipole to correspond to a velocity of $402 \pm 183$ km\,s$^{-1}$ which is consistent with our inferred motion through the CMB.

This paper is organised as follows. In \S~\ref{sec:kinematic} the kinematic dipole is defined while \S~\ref{sec:method} describes our methods for estimating the total dipole. In \S~\ref{sec:wise}, we introduce the dataset used and \S~\ref{sec:ONE} describes how we minimising contamination by foreground stars. In \S~\ref{sec:TWO} we describe methods to remove sources at low redshift in order to  reduce the clustering dipole and in \S~\ref{sec:residual} estimate the residual clustering dipole in our final selection, from theory as well as the $z=0$ halo catalogue of the Dark Sky simulations. Finally in \S~\ref{sec:result}, we estimate the velocity of the Solar system barycentre. 

\section{The Kinematic Dipole}
\label{sec:kinematic}

An observer moving with velocity $v$ in the rest frame of an intrinsically isotropic distribution of sources observes a dipolar modulation in the number count of sources with amplitude \citep{baldwin84, Itoh10}:
 
\begin{equation}
{\cal D}_{\rm kin}={v\over c}\,\left[2+x(1+ \alpha)\right]\,,
\label{eq:velocity-dipole}
\end{equation}
where $x$ and $\alpha$ are flux indices, defined through the integral source counts
\begin{equation}
{\rm d}N/{\rm d}\Omega (>S) = k S^{-x} ,
\label{eq:ellisbaldwin}
\end{equation}
and the flux density at a fixed observing frequency:
\begin{equation}
S_{\rm obs}=S_{\rm rest}\delta^{1+\alpha} .
\label{eq:alpha}
\end{equation}

\section{The dipole estimators}
\label{sec:method}

The dipole anisotropy in a catalogue can be estimated by calculating the difference in the number of sources in the upper and lower hemispheres. In order to estimate the dipole direction the direction of orientation of the hemispheres can be varied until the maximum hemispherical number count is obtained. The strength of the dipole is then given by 

\begin{equation}
d_{\rm HC} = \frac{N^{\rm UH}-N^{\rm LH}}{N^{\rm UH}+N^{\rm LH}} \times \hat{r}_{max}
\label{eq:hemisphericcount}
\end{equation}
where $N^{\rm UH}$ and $N^{\rm LH}$ are the numbers of sources in the upper and lower hemispheres respectively. We scan a healpy map of nside=32, to find the direction with the maximum hemispheric difference in number count \citep{colin17}.

While this estimator is robust its statistical variability is high and this produces a biased estimate of the magnitude of the dipole as shown in \citep{colin17}. Therefore we also use a 3-dimensional linear estimator \citep{crawford09}, defined by

\begin{equation}
d_{\rm 3D} = \frac{1}{N}\sum_{i=0}^{N} \hat{r}_i ,
\label{eq:crawfordesti}
\end{equation}

\noindent 
where $N$ is the number of sources in the catalogue and $\hat{r}_i$ is the unit vector in the direction of the source $i$. The statistical properties of this estimator have been studied previously \citep{rubart13,colin17}. 

The bias of these estimators is a measure of how the statistical noise due to finite sample size affects the measurement of the total dipole. It can be precisely quantified for a sample of a given size as was done in \citep{colin17} only when the total expected non-noise dipole is well known. In the presence of an additional component of the dipole such as ${\cal D}_{\rm cls}$, the bias factors estimated in \citep{colin17} effectively serve as upper limits to the total bias.

\section{WISE and AllWISE data}
\label{sec:wise}

The Wide-field Infrared Survey Explorer (WISE) mapped the sky in 2010 at 3.4, 4.6, 12, and 22 $\mu\,m$ (W1, W2, W3, W4) with an angular resolution of $6.1^{\prime\prime}, 6.4^{\prime\prime}, 6.5^{\prime\prime}, {\rm and}\, 12.0^{\prime\prime}$ respectively.
The WISE-allsky catalogue contains positions and the four-band photometry for 563,921,584 objects which include both stars and galaxies \citep{WISE}. 

The AllWISE catalogue supersedes the earlier WISE catalogue by combining data from the WISE cryogenic \citep{WISE} and NEOWISE \citep{NEOWISE} post-cryogenic survey phases and presently forms the most comprehensive view of the full mid-infrared sky. The AllWISE source catalogue contains astrometry and photometry for 747,634,026 objects \citep{cutri13}. In addition to increased depth, sensitivity and improved flux variability, the AllWISE Source Catalogue provides an estimate of the apparent motion of each source, exploiting the two independent WISE sky coverage epochs. 

\begin{figure}
\centering
\includegraphics[width=0.95\columnwidth]{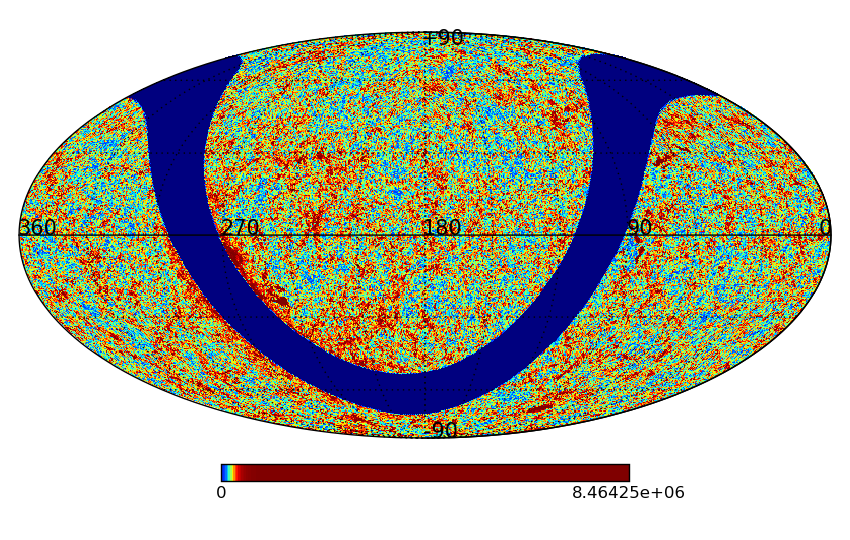} 
\caption{The density of the AllWISE catalogue [Number of sources/steradian] with the Galactic plane removed at $\pm 10\degree$ and the galaxy selection of \citet{szapudi15} applied, corresponding to catalogue {\sf\large(I)} of table \ref{tab:AllWISE}. This is a healpix map of nside 128.}
\label{fig:wise}
\end{figure}

\section{Data preparation I: Star-galaxy separation}
\label{sec:ONE}

The vast majority of sources in the WISE-allsky and AllWISE catalogues are stars within our own Galaxy. These have to be removed before a meaningful study of the dipole can be carried out. The star contamination can be estimated by cross-matching with SDSS \citep{SDSS} with a tolerance of  $1^{\prime\prime}$, as was previously done \citep{Yoon14}. As SDSS is a spectroscopic survey, it uniquely identifies stars and galaxies and hence serves as a control sample for star-galaxy separation. 

\subsection{Magnitude cuts and Galactic plane removal}
\label{sec:ONEONE}

\begin{figure*}
\includegraphics[width=0.9\columnwidth]{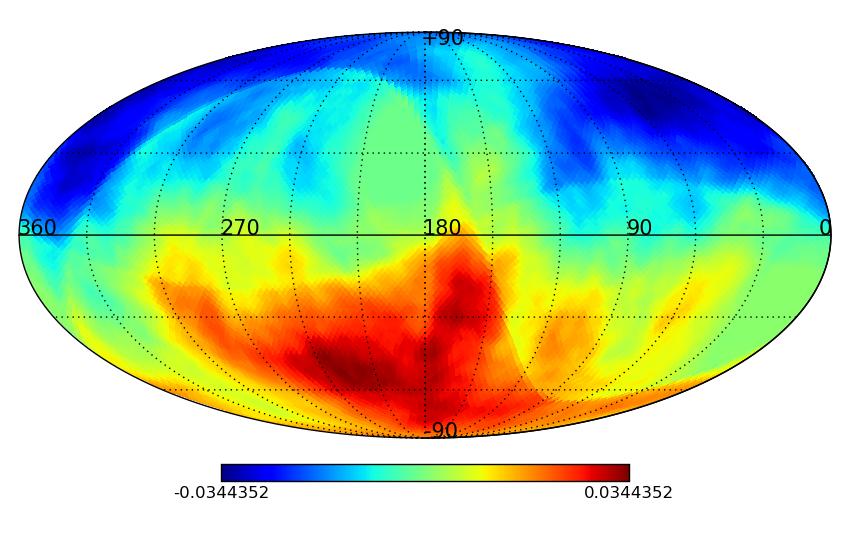} 
\includegraphics[width=0.9\columnwidth]{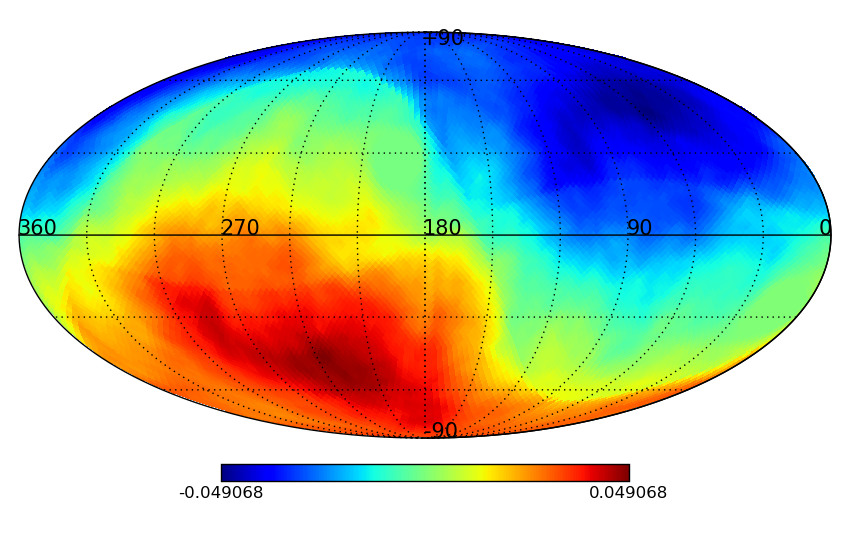} 
\caption{The left and right panels show the galaxy selection from WISE All-Sky and AllWISE catalogues, in equatorial coordinates. For each direction in the sky, the colour encodes: $(N^{\rm UH}-N^{\rm LH})/(N^{\rm UH}+N^{\rm LH})$, where $N^{\rm UH}$ and $N^{\rm LH}$ are the numbers of sources in the upper and lower hemisphere respectively. The directions are scanned using a healpy map of nside 32. The direction of the dipole is about $66\degree$ and $70\degree$ away from that given by the CMB for the WISE All Sky and AllWISE selections, while its magnitude is 3.4\% and 4.9\% respectively. These are healpix maps of nside 32.} 
\label{fig:dipolewiseallwise}
\end{figure*}

\indent
In order to remove stars contaminating our sample of galaxies, we follow \citet{szapudi15} who provide a separation strategy for objects in the WISE-allsky catalogue which have also been observed in 2MASS survey \citep{2MASS}. A sample of galaxies with 76\% galaxy completeness and just $\sim$2\% star contamination can be obtained by rejecting sources with $W1$ magnitude greater than 15.2 and $W1 - J_{\rm 2MASS} > -1.7$. Making a further $J_{\rm 2MASS} < 16.5$ cut \citep{Yoon14} and masking out contaminated regions in the WMAP Galactic dust mask can reduce the star contamination of the sample to 1.2\% (70.1\% galaxy completeness). However for the final sample to remain unbiased to dipole direction estimators, a cut removing sources at Galactic latitudes $|b| < 10\degree$ is preferred. After these cuts on the WISE-allsky catalogue we obtain a sample of  $\sim$2.359 million objects with a star-contamination of 1.8\% and 74\% galaxy completeness, while the AllWISE catalogue yields $\sim$2.367 million objects with a star contamination of 1.9\% and the same completeness.

The hemispherical number count estimator can now be applied to both the WISE-allsky and AllWISE samples to estimate the direction and magnitude of the maximum dipole anisotropy in the sky.

For the sample derived from WISE-allsky, we find that the dipole is in the direction RA=$228.6\degree$  and DEC=$-52.8\degree$ (or $l=323.67\degree$ and $b= 4.2175\degree$ in Galactic coordinates) and corresponds to a hemispherical number count difference of $3.4\%$. For the AllWISE catalogue we find a similar direction for the dipole of RA=$237.4\degree$ and DEC=$-46.6\degree$ ($l=331.9\degree$ and $b=6.02\degree$) and a hemispherical number count difference of $4.9\%$. These directions are significantly different from that of the CMB dipole, being $\sim66\degree$ and $\sim70\degree$ away respectively. Our results are in broad agreement with the previous findings e.g. by~\cite{Yoon14}, who also determined the galaxy bias $b$ at this stage to be $1.41 \pm 0.07$ (see Sec. \ref{sec:extflag} for more details). While a fully kinematic origin for these large dipoles would require velocities higher than $\sim$4000 km\,s$^{-1}$, the fact that their direction points to low Galactic latitudes is indicative of residual contamination by stars.

\subsection{Removing sources with large apparent motion}
\label{sec:pm}

\begin{figure}
\includegraphics[width=0.95\columnwidth]{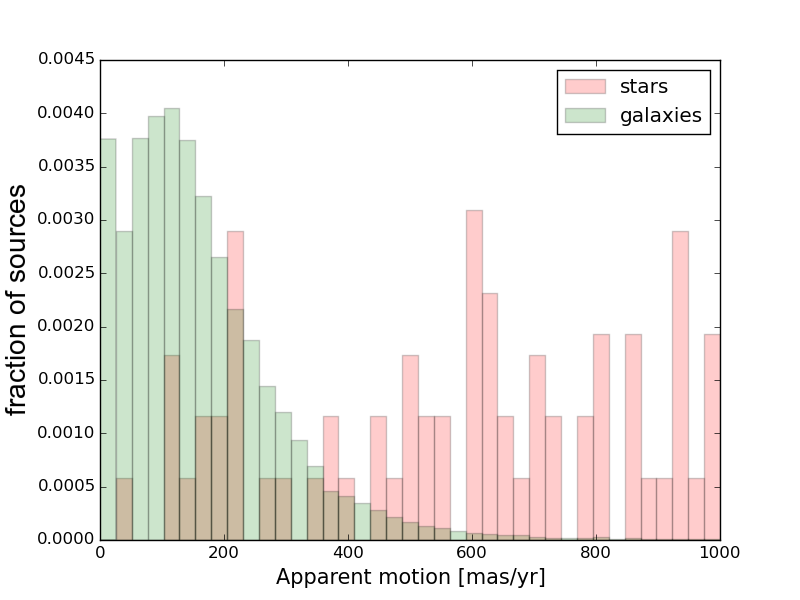}
\caption{The distribution of the AllWISE best-fit apparent motion for sources identified as galaxies, and those identified as stars through SDSS cross correlation. The normalisation of the stars have been scaled up (by a factor of $\sim$ 30) for visibility.}
\label{fig:apparentmotion}
\end{figure}

To further suppress star contamination in the catalogue, we widen the symmetric band around the Galactic plane within which sources are removed to $| b | < 15\degree$, leaving behind 2.09 million sources in AllWISE. Subsequently we use the apparent motion measurements as follows. Most objects in AllWISE have been observed only in two or three epochs, and consequently the proper motion and parallax components of the apparent motion cannot be disentangled. In general, it is the closest objects to the Sun that have substantial proper motions. These objects also have significant parallaxes. The stars that remain in our sample subsequent to the even wider Galactic latitude cuts at $\pm 15\degree$, reside at high Galactic latitudes and are hence nearby. Consequently, the apparent motion provides an excellent discriminator between extragalactic and Galactic objects\citep{vieira17}.

The value of the apparent motion, $am$, is given by the sum of the best fit proper motion in RA and DEC separately:

\begin{equation}
am=\sqrt{({\rm cos}\,({\rm DEC})\,\, \times {\rm RA}_{\rm am})^2+{\rm DEC}_{\rm am}^2}\,
\end{equation}

\noindent
where ${\rm RA_{am}}$ and ${\rm DEC_{am}}$ are the motions in RA and DEC calculated separately.
We combine these into one variable, the distribution of which is shown in Fig.~\ref{fig:apparentmotion}. Objects for which the motion fit fails to converge, which constitute $\sim 2.6\%$ of the sample are considered to have zero motion at this stage. \citet{Kurcz16} examine the quality of the apparent motion fits and conclude that they are reliable only for those sources for which the apparent motion fit has a signal-to-noise ratio larger than 1, albeit with a different definition of the apparent motion. The impact of applying both this more stringent criterion on the apparent motion, as well as assuming the objects with a failed motion fit to have zero motion, are re-examined at the final stage of our analysis (see \S \ref{sec:result}).

After removing all sources with $ am > 400$ mas/yr, we are left with about 1.91 million objects. Among the 36,086 sources that correlate to SDSS sources within $1^{\prime\prime}$ we find 34 stars, implying a star contamination of less than 0.1\%.

\section{Data preparation II: Suppressing the clustering dipole}
\label{sec:TWO}

\noindent The clustering dipole $({\cal D}_{\rm cls})$ is dominated by the contribution from nearby sources so should decrease at high redshifts. The kinematic dipole $({\cal D}_{\rm kin})$, due to the motion of the observer caused by the anisotropic distribution of mass seen in the clustering dipole, is however independent of the distance to the sources. 

To suppress the contribution of the local clustering dipole to the total dipole and extract the kinematic dipole, it is desirable to remove as many sources as possible at low redshifts, in a directionally unbiased manner. The various steps in the process of suppressing the clustering dipole are described in the following subsections.

WISE being a photometric instrument, the AllWISE catalogue does not provide redshift measurements. We estimate the redshift distribution of these data by cross matching with the Galaxy and Mass Assembly (GAMA) catalogue \citep{GAMA}. This is a spectroscopic survey of about 300,000 galaxies down to $r < 19.8$ magnitude over about 286 degree$^2$. The GAMA survey builds on the previous spectroscopic surveys such as the SDSS which we have used already to estimate the star contamination.

\begin{figure}
\includegraphics[width=\columnwidth]{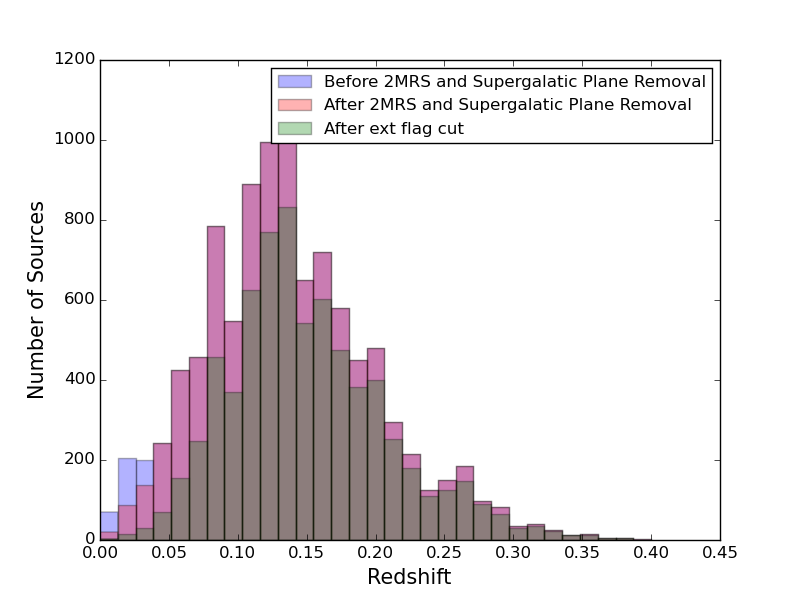} 
\caption{Redshift distribution for 5400 sources of AllWISE that are matched to those of GAMA survey. The median redshift is 0.137-0.164 depending on the masks.} 
\label{fig:gama}
\end{figure}

Of the 5,620 AllWISE sources at this stage that fall within the solid angle scanned by GAMA, 5,491 have cross-matched counterparts. The redshift distribution of these sources is shown in Figure~\ref{fig:gama} which also indicates how the distributions evolves in the later stages of this analysis. 

\subsection{Removing the supergalactic plane and sources correlating with 2MRS at $z<0.03$.}
\label{sec:2mrsSG}

A large fraction of the mass in the nearby universe, out to $z=0.03$, is known to be clustered along a planar structure known as the supergalactic plane. In order to exclude this, we add a supergalactic latitude cut of $\pm 5 \degree$ which ensures that most of the  local superclusters that lie on this plane are removed. Since both the galactic and the supergalactic planes form great circles in the celestial sphere, removing an area centered on them leaves the direction of the dipole estimators unbiased.

In order to further suppress any local super-structures that lie outside the supergalactic plane, we cross-correlate our AllWISE galaxy catalogue with the 2MRS catalogue \citep{huchra12} and remove all objects that are common to the two catalogues. This is done by identifying all AllWISE sources that are within $1^{\prime\prime}$ of 2MRS sources out to $z=0.03$, beyond which 2MRS is not complete. Of the 24,648 2MRS sources below redshift $z=0.03$, only 2,392 have AllWISE counterparts at this stage --- in contrast to  \S~\ref{sec:ONEONE}, when all 24,648 sources did have counterparts. Consequently, the impact of removing these sources is small.

Subsequent to these cuts we are left with $\sim$ 1.71 million objects. The median redshift at this stage was found to be $\sim 0.137$ and the 3D linear estimator of Eq. \ref{eq:crawfordesti} finds the direction and the magnitude of the dipole to be RA=$177.4\degree$, DEC=$-49.9\degree$ ($l=292.9\degree$, $b=11.7\degree$) and 0.017 respectively. The dipole direction is now $43.7\degree$ away from the CMB dipole. Evidently the removal of local structures slightly reduces the amplitude of the dipole (previous value was 0.018) and  brings its direction closer to that of the CMB. 

\begin{figure}
\includegraphics[width=\columnwidth]{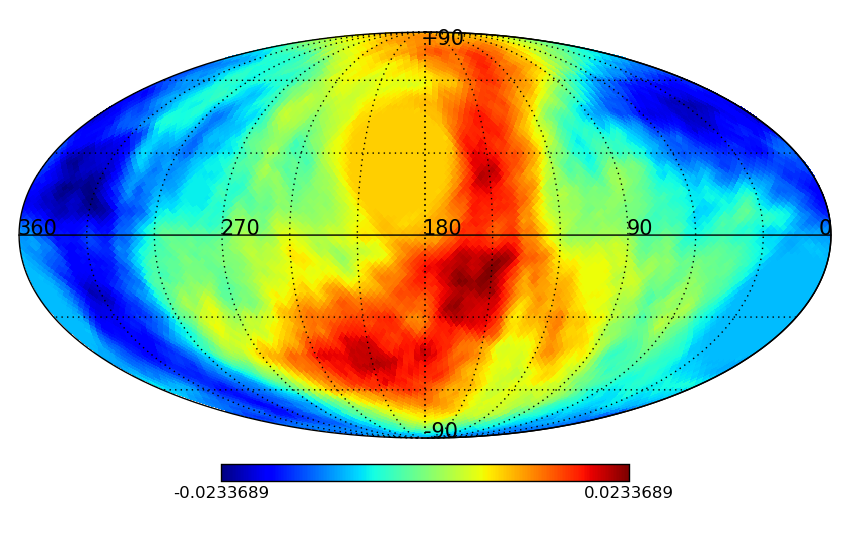} 
\caption{The hemispherical count map, in equatorial coordinates, of the AllWISE-galaxy selection as described in \S~\ref{sec:extflag}. This is a healpix map of nside 32.}
\label{fig:hcountfinal}
\end{figure}

\subsection{Discarding extended sources}
\label{sec:extflag}

The WISE satellite has an angular resolution of $\sim 6.1^{\prime \prime}$ in the 3.4 $\mu$m band, which corresponds to $2.96 \times 10^{-5}$ radians. Galaxies, which are typically a few tens of kpc across, are resolved as extended sources at distances less than a few hundred Mpcs. Galaxies of similar size at larger distances are contained within the angular beam size of the detector and appear to be point sources. Hence discarding extended sources at this stage can significantly suppress the fraction of nearby objects. The AllWISE catalogue provides a variable 'ext\_flg', which has a value of zero if the morphology of the source is consistent with the WISE point spread function, and not associated with a known 2MASS extended source. Higher values of the variable indicate high goodness of fits for extended source profiles. 

Therefore we select only sources with `ext\_flg=0', which leaves us with a sample of 1.23 million sources. The median redshift at this stage is found to have increased to 0.164, testifying to the suppression of low redshift sources.  Applying the 3D linear estimator of eq.(\ref{eq:crawfordesti}) to this sample, we find the dipole to be in the direction RA=$166.2\degree$, DEC=$-15.7\degree$ ($l=269.17\degree$, $b= 40.17\degree$), {\it i.e.} only $8.8\degree$ away from the CMB dipole. Its magnitude is now 0.0124, a significant reduction from the previous value of 0.017 (see \S~\ref{sec:2mrsSG}).  

\begin{figure}
\includegraphics[width=\columnwidth]{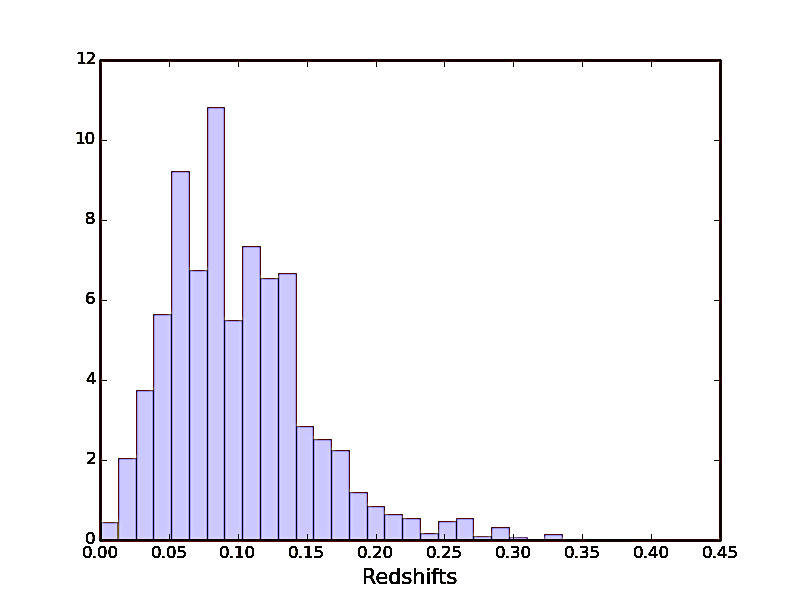} 
\caption{The redshift distribution of the sources (with ext\_flg$\neq$0) which are discarded in \S~\ref{sec:extflag}. The median redshift is 0.091. As explained in \S~\ref{sec:strucdipole}, these sources induce a velocity of at least 85 km\,s$^{-1}$ on the Local Group.}
\label{fig:redshiftdistroextended}
\end{figure}

If we further widen the Galactic plane cut to $\pm 20 \degree$, then the dipole direction swings to RA=$172.6\degree$, DEC=$-6.6\degree$ ($l=269.7\degree$, $b= 51.0\degree$), which is just $4.5\degree$ away from the CMB dipole, with a magnitude of 0.011 according to the 3D estimator. The hemispheric-count estimator of eq.(\ref{eq:hemisphericcount}) finds the dipole to lie towards RA=$151.9\degree$, DEC=$-15.7\degree$ ($l=255.1\degree$, $b= 31.5\degree$) which is $18.0\degree$ away from the CMB dipole, with a magnitude of 0.023 which is twice that found from the 3D estimator. The hemispherical count map at this stage is shown in figure \ref{fig:hcountfinal}. The discrepancy between the two estimators may be due to the larger bias of the hemispherical count estimator \citep{colin17}.

The galaxy bias, $b$ of this final sample can be evaluated using the method of \citet{szapudi15} and the SpICE~\citep{Szapudi:2000xj} software package and is found to be $1.27 \pm 0.12$ (excluding the $l=1$ mode from the fit). The lowering of the galaxy bias can be attributed to both the increase in the median redshift of the sample due to suppression of nearby objects, as well as the removal of extended galaxies, which correspond to the largest and most massive galaxies which are known to have a higher bias.

\subsection{Removal of individual local superclusters}
\label{sec:localsc}

\begin{figure*}
\includegraphics[width=\columnwidth]{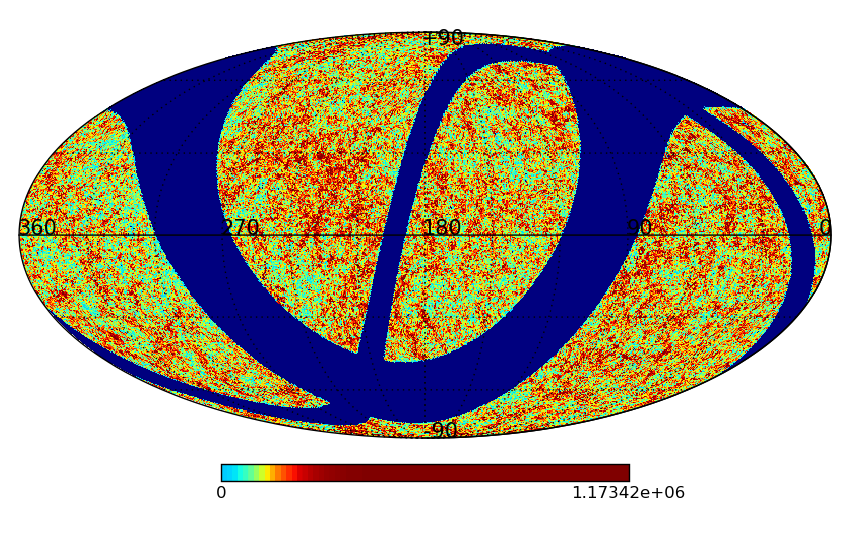} 
\includegraphics[width=\columnwidth]{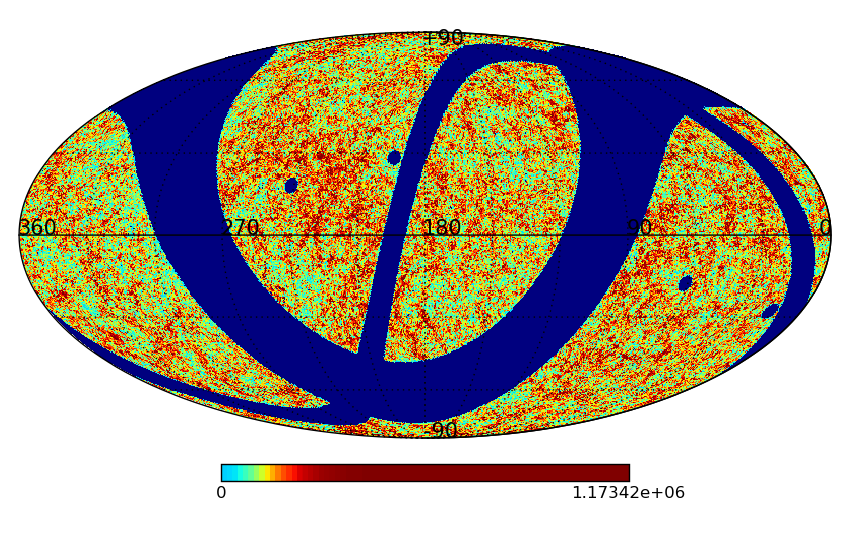} 
\caption{The densities of the final AllWISE-galaxy selections [Number of sources/steradian], corresponding to catalogues {\sf\large IV} (left) and {\sf\large V} (right) of table \ref{tab:AllWISE}. The punctures around the most prominent superclusters  of the local Universe, Shapley, Hercules and Coma (and their symmetric images), are performed as a robustness test. These are healpix maps of nside 128.}
\label{fig:punctured}
\end{figure*}

The positions of nearby clusters and superclusters are well-known. To test the robustness of the direction and magnitude of the total dipole estimated in \S~\ref{sec:extflag} with respect to contamination by these sources, we further mask out the sky around the Shapley supercluster and the Coma and Hercules Clusters. To keep the dipole estimators directionally unbiased, an equivalent mask has to be applied to the sky diametrically opposite to these directions (see Fig.~\ref{fig:punctured}).

The precise details of these ``punctures'' are as follows: $4\degree$ radius around Shapley at RA=$202.5\degree$,  DEC=$-31.0\degree$, $3\degree$ around Hercules at RA$=241.3\degree$, DEC$=17.75\degree$  and around Coma supercluster at RA$=194.95\degree$, DEC$= 27.98\degree$. After these cuts, there remain 1,193,188 sources. The results are presented in Table \ref{tab:AllWISE}. The removal of one or more of these sources affects the direction of the dipole by only $1-4\degree$ and the change in dipole magnitude is also insignificant. Most of these local structures have already been accounted for in the previous steps, hence these additional cuts have negligible impact as expected.

\section{The residual clustering dipole}
\label{sec:residual}

We can evaluate theoretically the expected clustering dipole in the AllWISE catalogue, from the angular power spectrum 
$C_l$ of the galaxy distribution. This is related to the three-dimensional spectrum $P(k)$ through (see {\it e.g.} \cite{huterer01,tegmark02})

\begin{equation}
C_l=b^2{2\over \pi}\int_0^\infty f_l (k)^2 P(k) k^2 dk\,.
\label{eq:cl}
\end{equation}
Here $b$ is the galaxy bias and the amplitude of the clustering dipole ${\cal D}_{\rm cls}$ is related to $C_1$, the $l=1$ mode of the angular power spectrum, as
$ {\cal D}_{\rm cls}=\sqrt {{9\over 4\pi} C_1} $. The filter function $f_l (k)$ is given by

\begin{equation}
f_l(k)=\int_0^\infty j_l(kr) f(r) dr\,,
\end{equation}
where $j_l$ is the spherical Bessel function which for the $l=1$ mode is ${\rm sin}(kr)/(kr)^2-{\rm cos}(kr)/(kr)$, $f(r)$ is the probability distribution for the comoving distance $r$ to a random galaxy in the survey which is given by the distributions in Figure~\ref{fig:gama}. For simplicity and lack of better information, we set the correction factor for $f(r)$ due to evolution to be unity (see {\it e.g.}~\cite{tegmark02}) and assume that $b$ is \emph{independent} of redshift. The function $f(r)$ is then proportional to

\begin{equation}
f(r) \sim {H(z)\over H_0 r_0}\, {dN\over dz}\,,
\label{eq:fr}
\end{equation}
which is normalised such that $\int _0^\infty f(r)dr=1$.

We set $r_0=c/H_0 \simeq 3000h^{-1}$ Mpc and use the code astropy~\citep{astropy} with cosmological parameters from Planck \citep{planck14} to obtain both the Hubble parameter as well as the comoving radius as a function of redshift, $r(z)$. The distribution $dN/dz$ can be approximated by splines fit to the different histograms of Figure \ref{fig:gama}. $P(k)$ can be obtained from CAMB \citep{CAMB}, at redshifts corresponding to the median redshift of the selection. Evaluating the above expressions numerically, we find the total average clustering dipole to be $\sim 0.0095\,b$ before the 2MRS-correlated and Supergalactic plane sources are removed, while after their removal it is $0.0068\,b$. It drops further to $\sim0.0052\,b$ after sources with ext\_flag$\neq 0$ are removed.

This evolution as we proceed through the various steps of selection towards our final sample, is in reasonable agreement with the findings reported in Table \ref{tab:AllWISE} (which also include the kinematic dipole contribution).

The theoretical approach employed so far refers to a typical observer in a $\Lambda$CDM universe. A more precise analysis should consider only galaxies similar to the Milky Way and its environment in a N-body simulation as we do below.

\subsection{Milky-Way like environments in the Dark-Sky simulations}
\label{sec:darksky}

We quantify the size and variance of the expected clustering dipole by looking at the $z=0$ snapshot halo catalogue of  `Dark Sky' --- a publicly available Hubble volume trillion-particle N-body simulation \citep{darksky}. Objects are sampled from the $z=0$ snapshot according to a redshift distribution which mimics that of the AllWISE galaxy selection. Subsequently, we use the 3D estimator (\ref{eq:crawfordesti}) to evaluate the dipole. Since we have not included the effect of aberration and Doppler boosting and calculate the angles from the coordinates of the objects in rest frame of the simulation, the dipole that we find is entirely \emph{non}-kinematical, {\it i.e.} purely due to the clustering effects.

The intrinsic dipole in the distribution of mass around an observer determines the magnitude and the direction of the velocity at that  position. Therefore to compare with our findings in the AllWISE galaxy selection, we examine catalogues constructed from the $z=0$ halo catalogue around observer halos similar to that of the Milky Way in mass and velocity, {\it viz.} with virial mass $M_{200}$ in the range $2.2 \times 10^{11}$ -- $1.4 \times 10^{12} M_\odot$ \citep{cautun14} and velocity in the range $600-650$ km\,s$^{-1}$. 

Each halo is assumed to correspond to a galaxy, and the most massive halos are selected in radial shells centered around the selected Milky Way-type observer in a directionally unbiased manner so as to produce a comoving distance distribution corresponding to the redshift distributions shown in Figure~\ref{fig:gama}. Such a catalogue is expected to have the same bias as the final AllWISE selection, although the galaxy-dark matter bias cannot be quantified from a simulation that includes only dark matter particles.

Subsequently, the 3D estimator (\ref{eq:crawfordesti}) is applied to such a catalogue to extract a purely intrinsic clustering dipole. 

\begin{figure*}
\includegraphics[width=0.9\columnwidth]{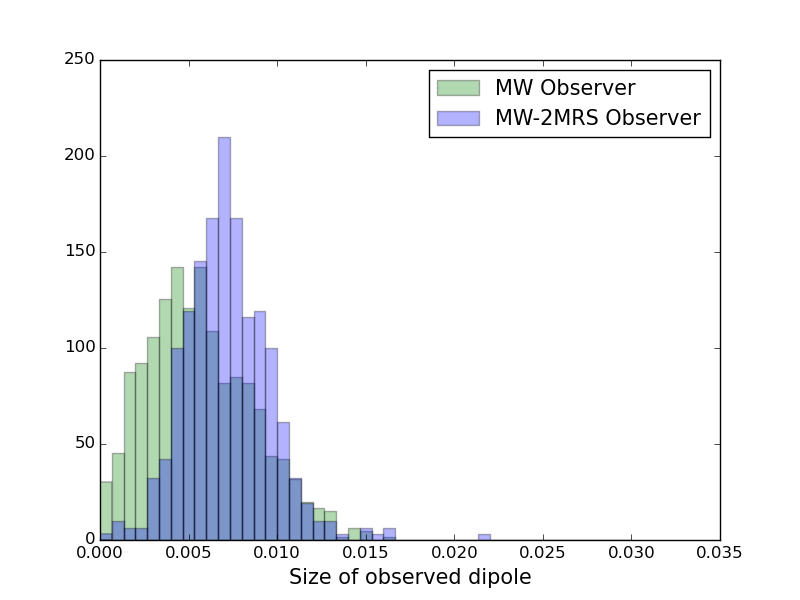} 
\includegraphics[width=0.9\columnwidth]{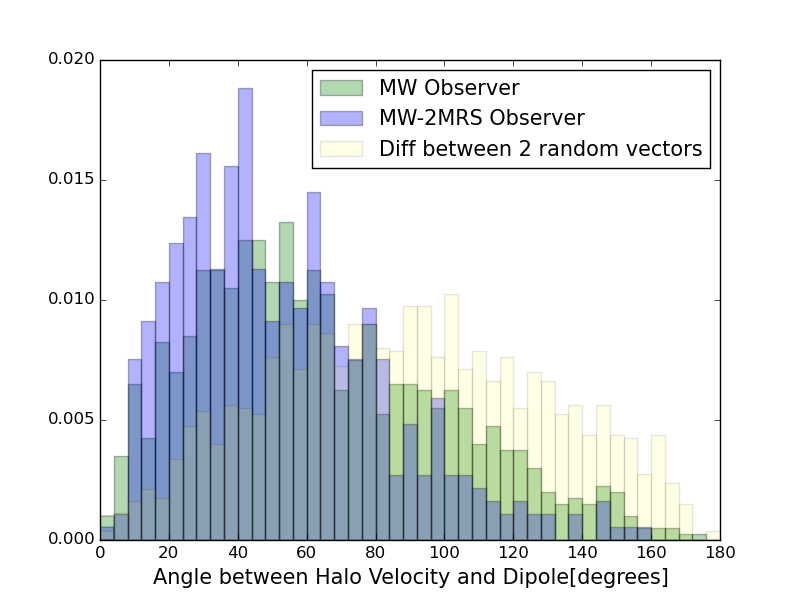} 
\caption{Left: Intrinsic clustering dipoles observed in 500 realisations of AllWISE-like galaxy catalogues from Milky Way-like halos (green) and Milky Way-like halos in an environment as found in 2MRS by \citet{lavaux10}, corresponding to a bulk flow velocity of the $z=0.03$ sphere in the range $240-280$ km\,s$^{-1}$ (blue). Right: The observed angle between the observer velocity and the observed clustering dipole direction. The distribution of angles expected between two isotropic random dipoles is shown (in yellow) for comparison.}
\label{fig:MW2MRSDipoledistr}
\end{figure*}

We assume that linear theory holds on these large scales, and that perturbations grow as $\delta(k, t_i) = \delta(k, t)D(t_i)/D(t)$. In order to accurately compare a catalogue constructed from a $z=0$ snapshot with an observed catalogue which is a light cone centred around the observer, the dipole contribution in each shell is scaled by the corresponding cosmological structure growth factor $D(z)/D(z=0)$. An examination of the catalogues constructed around 500 such observers yields dipoles with magnitudes and directions as shown in Fig.\ref{fig:MW2MRSDipoledistr}. 

To mimic our environment as closely as possible we now restrict ourselves to observers satisfying a  more stringent criterion. Previous work has shown that the velocity of the Local Group in a rest frame of radius of $\sim120$ Mpc (corresponding to $z \sim 0.03$) is $\sim 350$ km\,s$^{-1}$ (see {\it e.g.} \citep{lavaux10}), implying that a bulk flow of velocity $\sim 240$ km\,s$^{-1}$ persists beyond $z=0.03$ \citep{colin11,feindt13}.  Of the $\sim 23,800$ halos satisfying the previous criterion, only 500 are found to satisfy the criterion that the average velocity of all halos within a 120 Mpc sphere around it should be greater than 240 km\,s$^{-1}$, making the probability of such a system $\sim 0.021$. 15,973  Milky Way-like observer halos had to be examined before 500 were found with a similar bulk flow greater than 220 km\,s$^{-1}$, corresponding to probability of $\sim 0.031$, while for a threshold of 250 km\,s$^{-1}$, the probability is only $\sim 0.009$. 

To estimate the residual clustering dipole in our sample, we constrain ourselves to such observers, excluding also neighbourhoods with velocities higher than the allowed ranges to avoid biases from regions with unusually large bulk flows. We find that the residual clustering dipole has a value of  $0.0076 \pm 0.0022$ (see Fig.~ \ref{fig:MW2MRSDipoledistr}) for the range 240--280 km\,s$^{-1}$. The value is $0.0079 \pm 0.0017$ and $0.0071 \pm 0.0021$ for the ranges 250--290 km\,s$^{-1}$ and 220--260 km\,s$^{-1}$ respectively. The quoted uncertainties on these values correspond to 1$\sigma$.

\subsection{The clustering dipole and the velocity}
\label{sec:strucdipole}

While two dimensional catalogues such as the AllWISE-galaxy selection do not have redshift information, the velocity imposed by the dipole anisotropy on the local group of galaxies can still be estimated by employing the fluxes as proxies for the distance, assuming a narrow range of intrinsic luminosities for the sources under consideration. This idea was initially proposed by Gott and subsequently used for numerous surveys (see {\it e.g.} \cite{yahil86,maller03,erdogdu06,chodorowski11}).

The velocity-acceleration relation is \citep{peebles80}

\begin{equation}
{\vec v}={2\over 3}{ f(\Omega_{\rm m})\over  H_0 \Omega_{\rm m}} {\vec g}\,,
\end{equation}
where the acceleration $\vec g$ is given by
\begin{equation}
{\vec g}={G\over b} \sum_i {M_i\over r_i^2} {\hat{r}_i}
\end{equation}
Here $b$ is the bias between the mass and galaxy distribution, $M_i$ is the mass of galaxy $i$ and $r_i$ its distance from the observer, $H_0$ is the Hubble parameter at $z=0$ and $f (\Omega_{\rm m}) \sim \Omega_{\rm m}^{0.55}$, is the derivative of the growth factor with respect to the natural logarithm of the redshift. The above expression can be rewritten as

\begin{equation}
{\vec g}= {G\over b} \sum_i \, L_i\,\, {M_i\over L_i} {{\hat{r}_i} \over r_i^2} \, ,
\end{equation}
where $L_i$ is the apparent luminosity of galaxy $i$. Under the assumption of an universal mass-to-light ratio this writes

\begin{equation}
{\vec g} = {4\pi G\over b} {M\over L} \sum_i {L_i \over 4\pi r_i^2} {\hat{r}_i} = {4\pi G\over b} {M\over L}\sum_i \, S_i \hat{r_i}\,,
\end{equation}
where we have used the flux-luminosity relation $S_i=L_i/(4\pi r_i)$. The universal mass to light ratio can be evaluated for a given survey as \citep{peebles93}

\begin{equation}
{M\over L}={3\Omega_{\rm m}\over 8\pi G}{H_0^2\over j}\,,
\end{equation}
where the luminosity density $j$ is evaluated from the luminosity function $\Phi(L)$ of galaxies in a particular wavelength band using

\begin{equation}
j=\int_0^\infty L\Phi(L) dL
\end{equation}
However, most flux-limited catalogues have a lower flux cut-off which too needs be taken into account.
For the WISE catalogue the $2.4\mu$ luminosity density has been evaluated to be
$j_{{\rm wise}, 2\mu m}=3.8\times 10^8 L_{2.4\mu \odot}{{\rm Mpc}^{-3}}$ 
where the Solar luminosity $L_{2.4\mu \odot}=3.34 \times 10^{-8} {\rm Jy Mpc}^{2}$ \citep{lake17}. However, this was obtained by selective spectroscopy of a small subset of galaxies with median redshift $z \sim 0.35$. We correct this by a factor of $(1.35/1.14)^{0.8}$ in order to adapt the measurement to the median WISE redshift.

Putting all this together we have

\begin{equation}
{\vec v}= {f(\Omega_{\rm m})\over b}{H_0\over j}\sum_i S_i \hat{r_i}\,.
\label{eq:velocity-flux}
\end{equation}
We evaluate  $\sum_i S_i \hat{r_i}$ for different subsamples of the AllWISE galaxy selection and the corresponding induced velocity of the Local Group, given in Table.~\ref{tab:clustering-dipole}. While the final galaxy selection induces a velocity of just $\sim 50$ km\,s$^{-1}$ on the Local Group, only lower limits can be inferred for the contributions of nearby subsamples to this velocity as these are dominated by extended sources for which WISE photometry is significantly underestimated \citep{WISE}. These velocities serve as consistency checks for the analysis. However, they do not affect the conclusions drawn in the previous or following sections.

\section{Results and Discussion}
\label{sec:result}

The total observed dipole of the final sample of 1,233,920 galaxies, with star contamination less than 0.1\% and sources at low redshift suppressed as much as possible with WISE photometry, is found to be $\cal{D}$ = 0.0124 in the direction RA=$166.2\degree$, DEC=$-15.7\degree$ ($l=269.17\degree$, $b= 40.17\degree$), which is just $8.8\degree$ away from the CMB dipole. Of the total 1,323,200 objects that would have remained had those with an apparent motion above 400 mas/yr \emph{not} been removed, 88,121 are found to have a S/N ratio $> 1$. Discarding only these objects, we obtain a catalogue of 1,235,079 objects, with $\cal D$ = 0.0123 in the direction RA=$167.5\degree$, DEC=$-16.3\degree$ ($l=270.4\degree$, $b= 40.1\degree$),  {\it i.e.} $9.3\degree$ away from the CMB dipole. If the 32,576 sources for which the motion fit failed to converge are also discarded in addition to those with an apparent motion above 400 mas/yr, then the direction moves to RA=$156.4\degree$, DEC=$-5.4\degree$ ($l=250.1\degree$, $b= 42.0\degree$), {\it i.e.} $11.6\degree$ away from the CMB dipole, with $\cal D$ = 0.0132. Thus both the magnitude and the direction of the final observed dipole are reasonably robust with respect to the details of the apparent motion measurement, which is an essential ingredient of the process of suppressing the star contamination. They are also robust with respect to the removal of individual local superclusters as described in \S \ref{sec:localsc}.

Subtracting the best estimate of the residual clustering dipole, $|{\cal D}_{\rm cls}|$ = 0.0076 $\pm$ 0.0022 (\S~\ref{sec:darksky}) from the total observed dipole $|{\cal D}|$ = 0.0124 (\S~\ref{sec:extflag}), we obtain $|{\cal D} -{\cal D}_{\rm cls}| = 0.0048 \pm 0.0022$. A catalogue of this size (1.2 million sources) is also expected to  have a \emph{random} dipole of size $\sim 0.001$, implying $|{\cal D}_{\rm kin}| = 0.0048 \pm 0.0024$ if we do a scalar subtraction. While this subtraction ought to be done vectorially, the precise direction of the structure dipole in the local Universe is unknown. However the close alignment of the total dipole observed in data with the CMB dipole, despite  $|{\cal D}_{\rm cls}|$ being significantly larger than $|{\cal D}_{\rm kin}|$, suggests that the two are closely parallel. Hence the vector subtraction can be approximated with a scalar subtraction of the magnitudes.

It is straightforward to evaluate the flux power-law index $x$ in eq.(\ref{eq:ellisbaldwin}) for a given catalogue in a single frequency band. However for WISE and AllWISE, the initial cuts and the cuts applied for star-galaxy separation depend on magnitudes in different bands, hence the index changes between the different bands \citep{SETIPaper}. Since our galaxy selection is driven primarily by a $W1$ magnitude cut, we confine ourselves to the $W1$ band.

\begin{figure}
\includegraphics[width=0.95\columnwidth]{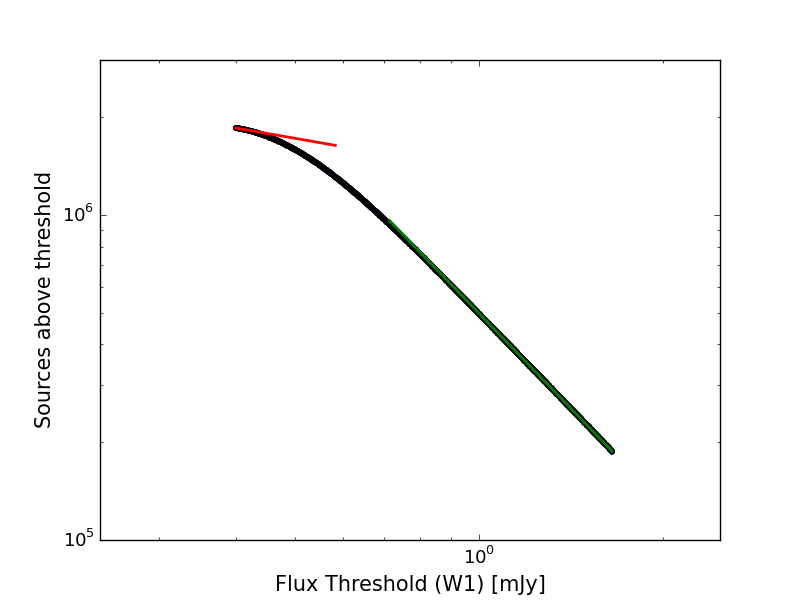} 
\caption{Variation of the AllWISE galaxy selection source count with the lower cut in flux. At the lower flux threshold, the best fit value (red line) of the power-law exponent $x$ in eq.(\ref{eq:ellisbaldwin}) is 0.75.}
\label{fig:nflux}
\end{figure}
The index of the flux function can be fitted from the data \citep{colin17}. The Doppler shift is more important for faint galaxies, hence the value of $x$ near the threshold is most relevant and is found to be 0.75 as shown in Figure~\ref{fig:nflux}. The spectral index  $\alpha$ (\ref{eq:alpha}) for galaxies in infrared depends on the classification of the galaxy. However, in the W1 band range, for most galaxy types, the spectral index varies between 0.8 and 1.0~\citep{SETIPaper}. Using a median spectral index of 0.9 and $x=0.75$ in eq.(\ref{eq:velocity-dipole}) yields a velocity of $420 \pm 213$ km\,s$^{-1}$ for the Solar system barycentre. The uncertainties include the statistical error in the clustering dipole, the error in the spectral index and the shot noise due to the random dipole from a finite sample, all added in quadrature.

\section{Summary}

The total observed dipole in the final AllWISE galaxy selection after suppressing star contamination and local source contribution is  $0.0124$ corresponding to a velocity of 1110 km\,s$^{-1}$ if interpreted as solely kinematic in origin.

While this seems anomalously high, theoretical expectations for a $\Lambda$CDM universe suggests that a clustering dipole of $\sim$ 0.006 is expected in a sample with the same redshift distribution as our final selection. This does not however account for our special local environment. To do so, we examine mock AllWISE galaxy selection-like catalogues generated from a $\Lambda$CDM Hubble volume simulation.  We search for haloes with velocities similar to that of the Milky Way embedded in an environment as observed in 2MRS with a bulk velocity of $\sim 240$ km\,s$^{-1}$ extending beyond $z=0.03$. We find that an intrinsic clustering dipole of size $0.0071 \pm 0.0022$ can arise for these observers. This lowers the inferred velocity of the Solar system barycentre to $430 \pm 213$ km\,s$^{-1}$, which is compatible with the value inferred from the CMB dipole. However, the estimate of the residual clustering dipole from theory is model dependent (here, a $\Lambda$CDM model with parameters fitted to Planck data) consequently the final value of the velocity \emph{cannot} be considered an independent measurement and serves only as a consistency test.

The structures in the redshift range 0.03--0.3 which give rise to such a large clustering dipole remain to be identified. In any case it is evident that we are \emph{not} typical observers --- the observed local velocity field is expected for only 2\% of Milky Way-like observers in the standard $\Lambda$CDM cosmology and this fraction drops rapidly as the local bulk velocity and its extent increases. The Six Degree Field Galaxy Redshift Survey (6dFGRS) extending out to $z \sim 0.05$ provides the most uniform galaxy peculiar velocity sample to date. If we adopt their estimate of the bulk velocity $397 \pm 68$ km\,s$^{-1}$ \citep{6dFGSv1,6dFGSv2}, then following the procedure in \S~\ref{sec:darksky} this fraction is found to be only 0.14\% for the median velocity, and 0.8\% for the $3 \sigma$ lower bound of 193 km\,s$^{-1}$. This provides a fresh perspective on interpretations drawn from analysis of cosmological data which assume that the observations are being made by a \emph{typical} observer.\\

\noindent
Note added: While this manuscript was being revised, an eprint by \citet{Hellwing:2018tiq}  appeared --- they too find that Local Group-like observers are rather untypical and explore the consequences for the interpretation of future cosmological data.

\begin{table*}
\caption{Observed dipoles in the galaxy selection described in \S~\ref{sec:ONEONE} with the hemispherical count estimator.}
\begin{tabular}{|c|c|c|c|c|c|c|c|c|}
 & & & & & & & &\\
 & & & & & & & &\\
{\large Catalogue} & $N$ & ${\rm RA}\degree$  & ${\rm DEC}\degree$  & $l\degree$ & $b\degree$ &  $\angle$ CMB &$|{\cal D}|$  \\
 \hline
 WISE All Sky     &   2,359,212  &  $228.6\degree$ & $-52.8\degree$ & $323.7\degree$ & $4.2\degree$ &  $66.9\degree$ & 0.034 \\
 AllWISE    &   2,367,162  &  $237.4\degree$ & $-46.6\degree$ & $331.9\degree$ & $6.0\degree$ &  $70.8\degree$ & 0.049 \\
\hline
\label{tab:WISE}
\end{tabular}
\end{table*}

\begin{table*}
\caption{Dipoles obtained with the 3D estimator in the AllWISE-galaxy selection as it evolves through the cuts described in \S~\ref{sec:ONE} and \S~\ref{sec:TWO}. Where multiple selection options are indicated, the final selection which carries over to the next stage is marked by a *. For reference, the CMB dipole has a magnitude of 0.0012.}
\begin{tabular}{|c|c|c|c|c|c|c|c|c|c|}
 & & & & & & & & &\\
{\large Catalogue} & $N$ & ${\rm RA}\degree$  & ${\rm DEC}\degree$  & $l\degree$ & $b\degree$ &  $\angle$ CMB &$|{\cal D}|$ &\\
AllWISE-2MASS & & & & & & & & &\\
\hline
{\sf\large (I)}\,(\S~\ref{sec:ONEONE});  &  &&&&&&& & \\
$|b| \geq  10\degree$ &2,367,162  &  $281.8\degree$ & $-57.7\degree$ & $338.1\degree$   & $-22.1\degree$     &  $96.4\degree$  & 0.032 &
\\
$|b| \geq  15\degree$ &2,092,276  &  $233.3\degree$ & $-64.7\degree$ & $319.2\degree$   & $-7.01\degree$     &  $73.3\degree$  & 0.021 & *
\\
 & & & & & & & & &\\
\hline
{\sf\large (II)}\, (\S~\ref{sec:pm});     & &          &          &             &             &         &     &   & \\
As in {\sf\large (I)} +& &           &          &        &             &         &        &  &\\
apparent motion $\leq$ 200 mas/yr  &   1,427,368 & $156.1\degree$  &  $-29.2\degree$  &  $268.2\degree$ &  $23.6\degree $   & $24.9\degree$           &    0.023   &  &\\
apparent motion $\leq$ 400 mas/yr  &   1,912,219 & $182.9\degree$  &  $-55.6\degree$  &  $297.3\degree$ &  $6.8\degree $   & $50.1\degree$           &    0.018   & * & \\
 & & & & & & & & &\\
 \hline
 {\sf \large (III)}\,(\S.~\ref{sec:2mrsSG});     &  &             &          &             &             &         &        &  &\\
As in {\sf\large (II)} +& &           &          &        &             &         &        & &\\
|supergalactic-$b|\geq 5\degree$ + sources& 1,718,619 & $177.4\degree$  &  $-49.9\degree$  &  $292.9\degree$ &  $11.7\degree $   & $43.7\degree$           &    0.017   & * &\\
 within $1^{\prime\prime}$  of
 2MRS sources,&  &          &          &        &             &         &     &    &\\
 with $z<0.03$ removed   &    &     &    &          &       &             &                  & &\\
 & & & & & & & & &\\
 \hline
 {\sf\large (IV)}\,(\S~\ref{sec:extflag}); &&&&&&&&&\\
 As in {\sf\large (III)} + sources & &           &          &        &             &         &        & &\\
  with ext\_flag $\neq$ 0 removed &1,233,920 &  $166.2\degree$  & $-15.7\degree$ & $269.17\degree$ & $40.17\degree$ & $8.8\degree$ &  0.0124 & * &\\
+${|b|} \geq 20\degree$ &1,084,178 &  $172.6\degree$  & $-6.6\degree$ & $269.7\degree$ & $51.0\degree$ & $4.5\degree$ &  0.011 &  \\
 &  &    &   & & & &  &  &\\
\hline
 {\sf\large (V)}\, (\S~\ref{sec:localsc});   & &           &          &        &             &         &      &   &\\
As in {\sf\large (IV)} +& &           &          &        &             &         &      &  & \\
 local superclusters removed  & 1,193,188 &    $174.3\degree$         &    $-11.1\degree$      &    $275.3\degree$         &       $47.7\degree$      &  $7.4\degree$  &  0.0119    & *  &\\
 \hline
\label{tab:AllWISE}
\end{tabular}
\end{table*}

\begin{table*}
\caption{Induced velocity from the clustering dipole in different subsamples. When dominated by extended sources, the WISE photometry significantly underestimates the flux and consequently only a lower limit is provided.}
\begin{tabular}{|c|c|c|c|c|c|c|}
 & & & & & & \\
 & & & & & & \\
{\large Catalogue} &  N  & ${\rm RA}\degree$  & ${\rm DEC}\degree$  &   $\angle$ CMB &$v$ (km\,s$^{-1}$)  \\
& & & & & & \\
\hline
AllWISE  - 2MRS (z< 0.03)  &&&&&   \\
 $\Delta b=\pm 5\degree$ &   24,648  &  $159.7\degree$ & $6.9\degree$  & $16.2\degree$  & $>116$ \\
 \hline
 AllWISE {\sf\large (III)} +    &     &   &   &   &   \\
 sources with ext\_flag $>$ 0 & 541,840 &$176.9\degree$ & $ -79.0\degree $ & $72.1\degree$ & $>85$ \\
 2MRS sources removed; $z<0.03$ &&&&& \\
 \hline
 AllWISE{\sf\large (IV)}     &&&&& \\
 $\Delta b=\pm 15\degree \Delta {\rm SGb}=\pm 5\degree$ &   1,358,233  &  $167.9\degree$ & $-40.9\degree$ & $33.9\degree$ & $\sim 39$ \\ 
 2MRS sources removed; $z<0.03$ &&&&& \\
 \hline
\label{tab:clustering-dipole}
\end{tabular}
\end{table*}

\section*{Acknowledgements}

MR acknowledges hospitality and support from the Institut d'Astrophysique Paris and SS acknowledges a DNRF Niels Bohr Professorship. We thank the anonymous referee for critical comments which improved the manuscript.


\end{document}